\documentclass{kluwer}    

\newdisplay{guess}{Conjecture}

\def\C{{\cal C}}
\def\beq{\begin{equation}}
\def\eeq{\end{equation}}
\def\F{{\cal F}}

\begin{document}                                                                                   
\begin{article}
\begin{opening}         
\title{Cosmology in a brane-universe} 
\author{David Langlois}  
\runningauthor{David Langlois}
\runningtitle{Cosmology in a brane-universe}
\institute{GReCO, Institut d'Astrophysique de Paris (CNRS)
\\
98 bis, Boulevard Arago, 75014 Paris, France}
\date{October 15, 2002}

\begin{abstract}
This contribution  presents the  cosmological 
models with {\it extra dimensions} that have been recently elaborated, 
which assume that  
ordinary matter is confined on a surface,  called {\it brane}, embedded in a 
higher dimensional spacetime. 

\end{abstract}

\end{opening}           

\section{Introduction}  
 The purpose of this contribution is to present a new approach to  
extra-dimensions, the  ``braneworld'' 
 scenario, where ordinary matter is trapped in a three-dimensional 
surface, called `brane',  embedded in a higher dimensional space.
 
This  idea must be contrasted with the traditional Kaluza-Klein 
treatment  of extra dimensions, where matter fields live {\it everywhere}
 in {\it compact} extra dimensions and 
 can  be described, via a  Fourier expansion, 
 as an infinite collection of 
 four-dimensional fields.  These so-called Kaluza-Klein modes
can be excited only when the energy at disposal exceeds their mass, typically
inversely proportional to the size, say $R$,  of the extra-dimensions. 
Presently, the energy scale reached in colliders is $E_{max}\sim 1$ TeV, 
which implies that 
\beq
R\lesssim (1{\rm TeV})^{-1}
\eeq
in the Kaluza-Klein picture.

In braneworlds, if only gravity propagates in the higher-dimensional 
spacetime, called {\it bulk},      the size of the extra-dimensions 
can be much larger than previously believed since four-dimensional gravity 
is only tested on scales above about a millimeter. 
Moreover, the 
four-dimensional Planck mass $M_p$ is in this context only a ``projection'' 
of the higher-dimensional (fundamental) Planck mass, which can be 
lower than $M_p$, thus offering a new perspective on the hierarchy 
problem  and suggesting the possibility 
that quantum gravity might be  closer than previously thought.

This new approach to dimensional reduction has been motivated by the 
latest developments in string/M-theory, in particular the emphasis  on branes
as loci where open strings end and define gauge fields,  as well
as the Horava-Witten model where gauge fields are defined on hyperplanes 
located at fixed points of the $Z_2$ orbifold symmetric eleventh dimension 
\cite{hw}. 

The cosmological consequences of these models have been studied in various 
ways.
The approaches tend to differ if one is a string theorist or  
a cosmologist. The former prefers to work with  models 
derived from string theory but often  too complex to tackle realistic 
cosmology  whereas the latter sacrifices  some aspects of the high energy 
phenomenology in order to get a tractable model. 
So far, the task  is too difficult to be satisfying from both viewpoints, 
but the hope  is that one can learn  from these two 
directions. Here, I focus on the cosmologist's strategy and present 
string-inspired, rather than string-derived, models, which consist of 
a self-gravitating brane-universe embedded in a five-dimensional  
bulk spacetime.

\section{The modified Friedmann equation}
The main motivation for exploring cosmology in models with extra-dimensions 
is that specific  signatures might be accessible 
{\it only} at very high energies, i.e. in the very early universe.  
One would thus like to investigate what kind 
of  relic imprints  could be left and  tested today via
 cosmological observations. This present contribution is devoted 
to   homogeneous brane cosmology and the reader is invited to refer to 
Nathalie Deruelle's  contribution  for a review on the important  
subject of cosmological perturbations in brane cosmology. 
   
In this section, I  describe the   cosmology 
 of a {\it self-gravitating brane-universe} 
embedded in an {\it empty} five-dimensional  
spacetime \cite{bdl99,bdel99,ftw99,sms99}. 
Assuming isotropy and homogeneity along three of its spatial dimensions 
(which correspond in the brane to our ordinary spatial dimensions) 
it is always possible to write the spacetime metric (at least locally 
in the vicinity of the brane) in the form
\beq
ds^2=- n(t,y)^2 dt^2+a(t,y)^2 \gamma_{ij}dx^idx^j+dy^2.
\label{metric}
\eeq
where $\gamma_{ij}$ is the maximally symmetric three-dimensional metric, 
with either negative, vanishing or positive  spatial curvature (respectively
labelled by $k=-1$, $0$ or $1$).

In these coordinates, our brane-universe is always located at $y=0$, and the 
cosmological scale factor for a brane observer is $a_0(t)\equiv a(t,0)$.
It is always possible to rescale the time coordinate so that it corresponds
on the brane to the usual cosmic time, i.e. $n_0(t)\equiv n(t,0)=1$.

The total energy-momentum tensor can be decomposed into a bulk 
energy-momentum tensor, which will be assumed to vanish here, 
 and a brane energy-momentum tensor, the latter 
being of the form
\beq
 T^A_{\, B}= S^A_{\, B}\delta (y)= \{\rho_b, P_b, P_b, P_b, 0\}\delta (y),
\eeq
where the  delta function expresses the localisation of   matter at the 
brane position $y=0$. The quantities $\rho_b$ and $P_b$ are respectively 
the total energy density and pressure in the brane and depend only
on time.
Allowing  for a cosmological 
constant in the bulk, $\Lambda$,   the 
five-dimensional Einstein equations read
\beq
G_{AB}+\Lambda g_{AB}=\kappa^2 T_{AB},
\label{einstein}
\eeq
where $\kappa^2$ is the gravitational coupling (and scales like the inverse 
of the cube of the fundamental mass in five dimensions). 

Instead of solving directly 
Einstein's equations with a distributional matter source, 
one can first obtain the general solution in the bulk and 
then apply boundary conditions at the brane location. 
The latter can be obtained from the 
 integration of Einstein's equations in the vicinity of the brane. 
According to these junction conditions,  
 the metric must be continuous and  the jump of the extrinsic curvature tensor 
$K_{AB}$  (related
to the derivatives of the metric with respect to $y$) depends on 
the distributional energy-momentum tensor, 
 \beq
\left[K^A_{\, B}
-K\delta ^A_{\, B}\right]=\kappa^2 S^A_{\, B},
\label{israel}
\eeq
where the brackets  denote the jump at the 
brane and the extrinsic curvature 
tensor is defined by 
$K_{AB}=h_{A}^C\nabla_C n_B$,
$n^A$ being the unit vector normal to the brane and $h_{AB}=g_{AB}-n_An_B$
the induced metric.

Assuming moreover that the brane is mirror symmetric, like in the 
Horava-Witten model,  
 the jump in the extrinsic curvature is twice its value on one 
side. 
Substituting  the ansatz metric (\ref{metric}) in (\ref{israel}), one ends up 
with the two junction conditions:
\beq
\left({n'\over n}\right)_{0^+}={\kappa^2\over 6}\left(3p_b+2\rho_b\right),
\qquad
\left({a'\over a}\right)_{0^+}=-{\kappa^2\over 6}\rho_b.
\label{junction}
\eeq
 
One can then solve  explicitly \cite{bdel99} 
 Einstein's  equations 
(\ref{einstein}) for the metric ansatz (\ref{metric}).
One finds in particular
that the geometry induced 
{\it in the brane} is governed by 
 the  equation 
\beq
H_0^2\equiv {\dot a_0^2\over a_0^2}={\kappa^4\over 36}\rho_b^2+{\Lambda\over 6}
-{k\over a_0^2}+{\C\over a_0^4}.
\label{fried}
\eeq
where  $\C$ is an integration constant.
This equation is analogous to the  (first)
Friedmann equation, since it relates the Hubble parameter to the 
energy density, but  is nevertheless different [the usual Friedmann equation
 reads $H^2=(8\pi G/3)\rho$]. 
 A striking property of this equation 
  is that the energy density of the brane enters 
{\it }quadratically on the right hand side in contrast with the standard 
four-dimensional Friedmann equation and its linear dependence on  
 the energy density. 
Another consequence of the five-dimensional Einstein equations 
(\ref{einstein}) is that 
the  energy conservation equation  
is unchanged 
and still reads
\beq
\dot\rho_b+3H(\rho_b+p_b)=0. 
\label{conserv}
\eeq

In the simplest case where $\Lambda=0$ and $\C=0$, 
one can easily solve the above cosmological equations 
(\ref{fried}-\ref{conserv}) for 
a perfect fluid with an equation of state $p_b=w\rho_b$ ($w$ constant).
One finds that the evolution of the scale factor is given by 
\beq
a_0(t)\propto t^{1\over 3(1+w)}.
\label{hesf}
\eeq
In the most interesting cases for cosmology, radiation and pressureless 
matter, one finds respectively  
$a\sim t^{1/4}$  (instead 
of the standard $a\sim t^{1/2}$) and $a\sim t^{1/3}$  (instead 
of  $a\sim t^{2/3}$).
Such behaviour  is problematic because it would spoil  
 nucleosynthesis. Indeed, the nucleosynthesis scenario depends on the 
balance between   the microphysical  reaction rates
 and the expansion rate of the universe, and changing the latter 
in a drastic way  between nucleosynthesis and today
 alters dramatically the predictions for the light element abundances.

The modified  Friedmann law (\ref{fried}), with the $\rho_b^2$ term
 but without the 
bulk cosmological constant (and without the $\C$ term) was  derived  
 just before  a new model describing a flat (Minkowski) world 
with one extra-dimension was proposed by Randall and Sundrum \cite{rs99b}. 
The new ingredient of this model 
was to endow  our brane-world  with a tension 
(constant energy 
density) and the five-dimensional bulk with a {\it negative} cosmological 
constant, the two being fine-tuned so that the effective four-dimensional 
Hubble parameter is zero in (\ref{fried}) (taking $\C=0$). It turns out that 
such a set-up gives the usual four-dimensional gravity, except on very small 
scales
 \cite{rs99b,gt99}.  

The recovery of ordinary gravity 
  suggested that  the  cosmological generalization of the Randall-Sundrum 
model    should  be compatible with standard cosmology 
at small energy  scales, as this was quickly verified \cite{cosmors1,cosmors2}.
Let us see how it works. In order to 
 go beyond the Minkowski geometry  and consider 
a non trivial   cosmology in the brane, one must  assume 
that the total energy density 
in the brane, $\rho_b$, consists of two parts, 
\beq
\rho_b=\sigma+\rho,
\eeq
the tension $\sigma$, constant in time, and the usual cosmological energy
density $\rho$. 
Substituting this decomposition into (\ref{fried}), one obtains 
\beq
H^2= \left({\kappa^4\over 36}\sigma^2+{\Lambda\over 6}\right)
+{\kappa^4\over 18}\sigma\rho
+{\kappa^4\over 36}\rho^2-{k\over a^2}+{\C\over a^4}.
\label{friedrs}
\eeq
Let us now fine-tune the brane tension and the bulk  cosmological constant so
that the  term between parentheses  vanishes (or at least is extremely small).
For $\rho \ll \sigma$, the next term dominates 
over the $\rho^2$  and
{\it one thus recovers the usual Friedmann equation at low energy}, 
with the identification
\beq
8\pi G= {\kappa^4\over 6}\sigma,
\label{newton}
\eeq
which also agrees with Newton's constant deduced from gravitational 
interaction between test masses. 
The third term on the right hand side of (\ref{friedrs}), 
quadratic in the energy density, 
provides a {\it high-energy correction} to the Friedmann equation 
which becomes significant when the value of the energy density approaches 
the value of the tension $\sigma$ and dominates  at higher 
energy densities. In the very high energy regime, $\rho\gg \sigma$, one 
thus recovers the unconventional behaviour of (\ref{hesf}) since the 
bulk cosmological constant becomes negligible.
 For an equation 
of state $p=w\rho$, with $w$ constant,  the 
conservation equation (\ref{conserv}) gives  as usual
\beq
\rho=\rho_* a^{-q}, \quad q\equiv 3(w+1),
\eeq
which after substitution in the Friedmann equation (\ref{friedrs}) yields
(for $k=0$ and $\C=0$)
\beq 
a(t)=\left[q m_* t\left(1+{q\over 2}\mu t\right)\right]^{1/q}, 
\eeq 
where we have introduced the two mass scales 
\beq
m_*\equiv  {\kappa^2\over 6}\rho_*, \qquad \mu\equiv \sqrt{-\Lambda/6}.
\eeq
 One sees that the evolution of the scale 
factor interpolates between the low energy regime and the 
high energy regime and that the transition time is of the order of $\mu^{-1}$, 
which is the characteristic  scale associated with the cosmological 
constant.

Finally, the last term on  the right hand side of (\ref{friedrs}) 
behaves like radiation and 
arises from the integration constant $\C$. This 
constant $\C$ is  analogous to the Schwarzschild mass, as we will see in 
the next section,  and it is  
related to the bulk Weyl tensor, which vanishes when $\C=0$. In a 
cosmological context, this term is constrained to be small enough 
at the time of nucleosynthesis in order to satisfy the constraints on the 
number of extra light degrees of freedom. 

 The metric  outside the brane can be also determined explicitly
\cite{bdel99}.
In the special  case $\C=0$, the metric has a very simple form and 
its components are given by   
\begin{eqnarray}
a(t,y)&=& a_0(t)\left(\cosh\mu y-\eta \sinh\mu|y|\right)\\
n(t,y)&=& \cosh\mu y-\tilde\eta \sinh\mu|y|
\label{bulk_metric}
\end{eqnarray}
where
\beq
\eta=1+{\rho\over\sigma}, \qquad \tilde\eta=\eta+{\dot\eta\over H_0}.
\eeq
The Randall-Sundrum model  corresponds to 
 $\rho=0$, i.e. $\rho_b=\sigma$, which 
implies  $\eta=\tilde\eta=1$ and 
$a(t,y)=a_0\exp(-\mu|y|)$.

As explained above, the Randall-Sundrum  version of brane cosmology
gives at sufficiently late times a cosmological evolution identical 
to the usual one. The model is thus viable if  
the low-energy regime  encompasses the periods that are well constrained
by cosmological observations. 
This essentially means  that 
nucleosynthesis must take place in the low-energy regime. 
This is the case if 
the energy scale associated with the tension is higher than 
the nucleosynthesis
energy scale, i.e.
\beq
\sigma^{1/4} \gtrsim 1 \ {\rm MeV}.
\eeq
Combining this with (\ref{newton}) implies for the fundamental mass scale
(defined by $\kappa^2=M^{-3}$)  the constraint 
$M \gtrsim 10^4 \ {\rm GeV}$. 
There is however a more stringent 
 constraint:
the requirement to recover ordinary gravity down to scales of the 
submillimeter order, which have been probed by gravity 
experiments \cite{grav_exp}. This implies
\beq
\ell=\mu^{-1} \lesssim  10^{-1} \ {\rm mm},
\eeq
which yields the constraint
\beq
 M \gtrsim 10^8 \ {\rm GeV}.
\eeq
Another parameter of the model is the  Weyl parameter $\C$. 
As mentioned above, its value is restricted by 
the  bounds on the number of additional 
relativistic degrees of freedom allowed during nucleosynthesis (usually 
expressed as the number of additional light neutrino species). 
Typically, this gives the constraint 
\beq
{\rho_{Weyl}\over \rho_{rad}}\equiv {\C\sigma\over 2 a^4\mu^2\rho} \lesssim
10 \%.
\eeq

\section{Moving brane in a static bulk}
In the previous section, we considered a specific system of coordinates such 
that the brane is always at $y=0$ and the metric is of the form (\ref{metric}).
Although this choice is very convenient from the  point of view of the brane, 
it does not give the simplest description of the bulk geometry. 
Indeed, it turns out that the required `cosmological 
symmetries' are so strong that the  geometry of the empty bulk is necessarily 
static, and in an appropriate coordinate system, is described by 
a metric of the form \cite{kraus99} 
\beq
ds^2=-f(R)\, dT^2+{dR^2\over f(R)}+R^2\, \gamma_{ij}\, dx^i dx^j, 
\label{adsmetric}
\eeq
where 
\beq
f(R)\equiv k-{\Lambda\over 6} R^2-{\C\over R^2}.
\label{f}
\eeq
The above metric is known as the five-dimensional Schwarzschild-Anti de 
Sitter (Sch-AdS) metric 
(with $\Lambda<0$). 
It is clear from (\ref{f}) that $\C$, as noted before, is the five-dimensional 
analog of the Schwarschild mass (the $R^{-2}$ dependence
 instead of the usual $R^{-1}$ is simply due to the different
 dimension of spacetime).

It can be shown explicitly, by considering  the appropriate coordinate 
transformation, that  the  manifestly static metric (\ref{adsmetric}) indeed 
coincides with the previous expression (\ref{bulk_metric}) for the metric. 
The simplicity of the metric (\ref{adsmetric}) is however counterbalanced 
by the fact that {\it the cosmological brane is necessarily moving} 
with respect 
to this coordinate frame.

The trajectory of the brane can be defined 
by its coordinates $T(\tau)$ and $R(\tau)$ given as functions of a 
parameter $\tau$. Choosing $\tau$  to be the proper time 
imposes the condition 
\beq
g_{ab}u^a u^b=- f\dot T^2+{\dot R^2\over f}=-1, 
\label{normalization}
\eeq
where $u^a=(\dot T, \dot R)$ is the brane velocity  and a dot 
stands for  a derivative with respect to the parameter $\tau$.
Using this normalization condition (\ref{normalization}), one finds that 
the components of the unit normal vector (defined such that $n_au^a=0$ and 
$n_an^a=1$) are given, up to a sign ambiguity, by
$n_a=\left(\dot R, -{\sqrt{f+\dot R^2}/f}\right)$.
The four-dimensional metric induced in the brane worldsheet is then 
directly given by 
\beq
ds^2=-d\tau^2+R(\tau)^2 d\Omega_k^2,
\eeq
which shows clearly 
that the brane scale factor, denoted $a_0$ previously, 
can be identified with the radial coordinate of the brane $R(\tau)$.

The dynamics of the brane is then obtained by writing the junction conditions
for the brane.
The `orthogonal' part of the junction conditions yields
\beq
K^i_{j}={\sqrt{f+\dot R^2}\over R} \delta^i_{j}={\kappa^2\over 6}\rho_b,
\label{junction_ij}
\eeq
which, after  substituting (\ref{f}) and 
rearranging, gives  exactly the 
 Friedmann equation (\ref{fried}) obtained before.  
There is also information  in the `longitudinal'
 part of the junction conditions, which can be rewritten as
  the standard  conservation 
equation (\ref{conserv}).
This confirms the complete equivalence between the `brane-based' and the 
`bulk-based' pictures. 

Let us add that the metric (\ref{adsmetric}) describes in principle only one 
side of the brane. In the case of a mirror symmetric brane, as assumed above,
 the complete spacetime is obtained by gluing, along the brane worldsheet,
 two copies of a portion of Sch-AdS spacetime. 
For an asymmetric brane, one can glue two
(compatible) portions of different Sch-AdS spacetimes. This can also be 
generalized to a system of several (`parallel') branes, which can 
be moving with respect to each other, thus suggesting the possibility 
of collisions \cite{lmw01}. This idea has recently 
attracted a lot of attention with the proposal that the cosmological 
Big Bang might be in fact such a brane  collision.
 
\section{Brane radiating gravitons into the bulk}
The analysis of the cosmological behaviour of the brane presented 
in the previous sections was based on the assumption of perfect homogeneity 
and isotropy along the three ordinary dimensions. In real cosmology, these 
symmetries hold only on average and there exist 
fluctuations on small scales, which create gravitational waves that can 
escape into the bulk. A consequence of this process is that the Weyl parameter
$\C$ is no longer constant \cite{hm01}. 

As I will now summarize, it is possible to build a simplified model 
that treats {\it self-consistently} the emission 
of gravitons, the backreaction on  the bulk geometry and the
 motion, i.e. cosmology, of  our brane-universe \cite{lsr02}. 
This model is based on the simplifying assumption that all gravitons 
are emitted only in the radial direction, i.e. perpendicularly to the brane.
The corresponding bulk energy-momentum tensor is thus of  the form
\beq
T_{AB}=\F k_A k_B,
\label{Tbulk}
\eeq
where $k^A$ is an ingoing null vector, which can be normalized 
so that  $k_A u^A=1$, where $u^A$ is the brane velocity vector.
The solution of the bulk Einstein equations is then the 
generalization  to  five dimensions of  
Vaidya's metric, 
which   describes the spacetime 
surrounding a radiating star. The associated metric
 reads
\beq
ds^2=-f\left(r,\,v\right)\,dv^2+2\,dr\,dv+r^2\,\delta_{ij}dx^idx^j, \quad 
f(r,v)=\mu^2 r^2-{\C(v)\over r^2}.
\label{vaidya}
\eeq
If $\C$ does not depend on $v$, the above metric is 
simply a rewriting of the Sch-AdS metric (\ref{adsmetric}) 
in terms of  the null coordinate 
$v=T+\int dr/f(r)$.  Note that, strictly speaking, the Vaidya 
spacetime corresponds
to an {\it outgoing} radiation flow, whereas in our case, we are interested 
in an {\it ingoing} radial flow because the brane, which emits the radiation, 
is in some sense located at the largest radius of spacetime (when 
getting away from the brane the radius, or scale factor, decreases).

Einstein's equations relate the energy flux $\F$ to the variation  of the
Weyl parameter, according to the expression
\beq
{d\C\over dv}={2\kappa^2\F\over 3}r^3
\left(\dot r-\sqrt{f+\dot r^2}\right)^2.
\label{sigma}
\eeq
The `orthogonal' junction conditions for the brane yield the same expression
as in (\ref{junction_ij}) and thus 
 the same brane Friedmann equation as before with the important 
change that the Weyl parameter $\C$ now depends on time.
The `longitudinal' junction conditions can be expressed as 
\beq
\dot \rho_b+3{\dot r\over r} (\rho_b+p_b)=-2\F,
\label{nonconserv}
\eeq
which differs from the previous conservation law (\ref{conserv})
by the nonzero right hand side which 
 represents the  energy loss, from the brane point 
of view, due to the escaping gravitons.

In order to close the system, one must evaluate the rate of graviton 
production in terms of the brane parameters. In the radiation era, one 
can show that the energy density loss rate $\F$ is proportional to 
$T^8$, so that one can write
\beq
\F={\alpha\over 12}\kappa^2\rho^2,
\label{F}
\eeq
where $\alpha$  depends on the number 
of relativistic degrees of freedom. $\alpha\simeq 0.019$ 
if all degrees of freedom of the 
standard model are relativistic.

One can now solve the coupled system consisting of (\ref{sigma}), 
(\ref{nonconserv}) with (\ref{F}), and the Friedmann equation. 
The high energy regime is  characterized 
by a rapid growth of the Weyl parameter due to an
abundant production of bulk gravitons. In the low energy 
 radiation era, the Weyl parameter  approaches a constant value, which
 means that  the production of bulk gravitons becomes negligible. 

This asymptotic value for $\C$ can be estimated analytically. If the 
present description is  valid   deep enough in the high energy regime,   
one finds 
\beq\label{rhoweyl}
\epsilon_W\equiv {\rho_{Weyl}\over \rho_{rad}}\simeq 2\times 10^{-3}
\eeq
at the time of nucleosynthesis (with the value of $\alpha$ given above).
This result must be compared with the present bound on 
additional  relativistic 
degrees of freedom allowed during  nucleosynthesis, which gives the 
constraint $\epsilon_W\lesssim 8 \times 10^{-2}$.

\section{Bulk scalar field}
Although brane cosmology has been mostly studied, out of simplicity, 
 for an  {\it empty} bulk, i.e. with only gravity 
propagating in the bulk, string models have prompted  the analysis 
of brane cosmology with bulk fields, the simplest example being a 
bulk scalar field. Such a model can be described by the action 
\beq
{\cal S} = \int d^5 x \,\sqrt{-g}\,\left[ 
\,{{}^{(5)}R\over 2\kappa^2} - \frac{1}{2} \,(\partial\phi)^2 
 - V(\phi)\,\right] 
+ \int_{brane} d^4 x \, {L}_{m}[\varphi_m,\tilde h_{\mu\nu}],
\label{action}
\eeq
where it is assumed that the four-dimensional metric $\tilde h_{\mu\nu}$, 
minimally 
coupled to the four-dimensional matter fields $\varphi_m$ in the 
brane, is conformally related to the induced metric $h_{\mu\nu}$, i.e.
\beq
\tilde h_{\mu\nu}=e^{2\xi(\phi)}h_{\mu\nu}.
\eeq
Variation of the action (\ref{action}) with respect to the metric
yields 
the five-dimensional Einstein equations (\ref{einstein}), 
where, in addition to the 
(distributional) brane energy-momentum tensor, there is now the scalar field
energy-momentum tensor. 
Variation of (\ref{action}) with respect to $\phi$ 
yields the equation 
of motion for the scalar field, 
which is of the Klein-Gordon type with  
a distributional source term since  the scalar field is coupled to the brane
via $\tilde h_{\mu\nu}$.
This implies that there is another junction condition at the brane location, 
now involving  
 the scalar field and  which is of the form
\beq
\left[n^A\partial_A\phi\right]=-\xi' T,
\label{junction_phi}
\eeq
where $T=-\rho+3P$ is the trace of the energy-momentum tensor (defined with
respect to $h_{\mu\nu}$).

Although the dynamics of the full system is very complicated in general, 
one can find analytical solutions in some cases, in particular with 
an exponential potential \cite{cr99,lr01,charmousis01}, 
\beq
V(\phi)=V_0\exp\left(-{2\over \sqrt{3}}
\lambda \kappa\phi\right).
\label{potential}
\eeq
For example, there exists a simple class of  static solutions,
described by the  metric  
\beq
ds^2=-h(R)dT^2+{R^{2\lambda^2}\over h(R)}dR^2+R^2 d{\vec x}^2,
\label{dil_stat}
\eeq
with 
\beq
h(R)=-{\kappa^2 V_0 /6\over 1-(\lambda^2/4)}R^2-\C R^{\lambda^2-2},
\eeq
where $\C$ is an arbitrary constant, 
and the scalar field 
\beq
{\kappa\over \sqrt{3}}\phi=\lambda\ln(R).
\eeq
To include a brane in this configuration, one must ensure that 
the three junction conditions, two  for the 
metric and one for the scalar field, are satisfied. It can be shown 
that  these three junction conditions are equivalent to the following 
three relations
\begin{itemize}  
\item  a generalized  Friedmann equation,
\beq
H^2={\kappa^2\over 36}\rho^2- {h(R)\over R^{2+2\lambda^2}}=
{\kappa^2\over 36}\rho^2+{\kappa^2 V_0 /6\over 1-(\lambda^2/4)}R^{-2\lambda^2}
+\C R^{-4-\lambda^2},
\eeq
\item a (non-) conservation equation for the energy density,
\beq
\dot\rho+3H(\rho+p)=(1-3w)\xi' \rho \dot \phi,
\eeq
\item 
a constraint on  the brane matter equation of state, which must be related 
to the conformal coupling according to the expression 
\beq
3w-1={\kappa\over\sqrt{3}}{\lambda\over \xi'}.
\eeq
\end{itemize}
The staticity of the bulk thus allows only a very contrived 
 cosmology in the brane.

In contrast with the empty bulk case, where the required symmetries impose
the bulk to be static, one can now find solutions with a non static bulk.
For example, with the same potential (\ref{potential}), 
a solution of the Einstein/Klein-Gordon bulk equations is given 
 by
the metric \cite{lr01}
\beq
ds^2={18\over \kappa^2 V_0}e^{4\lambda^2T} \, e^{4\lambda\sqrt{\lambda^2- 1}
 R}
\left(-dT^2+dR^2\right) + e^{4T} d{\vec x}^2,
\eeq
and the scalar field configuration 
\beq
\phi=4\left(\lambda^2 T +\lambda\sqrt{\lambda^2- 1}\,  R\right).
\eeq
It is possible to embed a (mirror) symmetric brane in this spacetime, with 
an equation of state $p_b=w\rho_b$ ($w$ constant). This  leads, 
 for a brane observer in the Einstein frame, to a cosmology with  
a  power-law expansion, but once more rather contrived.

\theendnotes

\end{article}
\end{document}